# Perla: A Conversational Agent for Depression Screening in Digital Ecosystems. Design, Implementation and Validation

Raúl Arrabales

Psicobótica Labs, Madrid, Spain
`raul@psicobotica.com`

**Abstract.** Most depression assessment tools are based on self-report questionnaires, such as the Patient Health Questionnaire (PHQ-9). These psychometric instruments can be easily adapted to an online setting by means of electronic forms. However, this approach lacks the interacting and engaging features of modern digital environments. With the aim of making depression screening more available, attractive and effective, we developed Perla, a conversational agent able to perform an interview based on the PHQ-9. We also conducted a validation study in which we compared the results obtained by the traditional self-report questionnaire with Perla's automated interview. Analyzing the results from this study we draw two significant conclusions: firstly, Perla is much preferred by Internet users, achieving more than 2.5 times more reach than a traditional form-based questionnaire; secondly, her psychometric properties (Cronbach's alpha of 0.81, sensitivity of 96% and specificity of 90%) are excellent and comparable to the traditional well-established depression screening questionnaires.

## 1 Introduction

Epidemiological studies show that depression and related mood disorders are amongst the most prevalent mental health problems in the world. The suffering of depression is not only a substantial personal burden but a global socioeconomic problem involving unhealthy family functioning, disability, absenteeism, loss of productivity and diminished social welfare. Lifetime prevalence of depression is between 10% and 15% for general populations, increasing their risk of suicide by a factor of 20 in relation to non-depressed population [1]. The prevalence of depression reaches much higher ratios when the focus is put in comorbidity with other mental health disorders. For instance, one large cohort study showed that more than 80% of persons with a current anxiety disorder also had a lifetime depressive disorder [2].

Given this situation, numerous initiatives have been implemented over time in order to improve the early detection of mood disorders, both in general populations and specific risk groups [3,4,5]. The context of these early detection projects has been traditionally located in primary care and specialized healthcare units, where mental health professionals are present and have direct access to patients, making it possible to conduct in-person assessments.

Typical settings of depression early diagnosis in secondary and tertiary care are very specialized, involving the use of scarce and expensive resources, such as trained psychologists, psychiatrists or other healthcare professionals and specific neuropsychological, behavioral or in-person interview-based assessment tools [6,7]. By contrast, the situation of early diagnosis of depression by primary care providers is much different and traditionally considered to be deficient. Old studies from the 80's and 90's considered that primary care physicians often failed to recognize depression symptoms [8,9]. Nowadays, depression misdiagnose and undertreatment in primary care is still a concern [10], and primary care physicians are encouraged to use screening tools, such as Patient Health Questionaire-2 (PHQ-2) [11], Patient Health Questionaire-9 (PHQ-9) [12] and Beck Depression Inventory (BDI) [13], so that appropriate treatment can be promptly initiated.

On one hand, we might consider that it is never too late to identify a depression case and to initiate the appropriate treatment, however, on the other hand, there is no doubt that early detection is of great value, even if only to reduce the time that the person is suffering. Additionally, there are other significant advantages to early detection of depression [14]. Some of the benefits of early diagnosis and intervention include the reduction of recurrent episodes and relapses [15], increase in social function and productivity, decreased absenteeism and better chance of remission [16].

Typically, the majority of patients with depression seek help in primary care for many reasons [16], some misperceive their symptoms as non-psychological in origin, many others are ashamed of their supposed sign of "mental weakness" and are reluctant to seek out help from a mental health service. Nowadays, in the era of digital ecosystems, in which we enjoy multitude of thriving and vibrant new ways of communicating over the Internet, the general population is increasingly turning to platforms like social media in search for mental health advise [17]. There is much debate whether this phenomenon is beneficial or dangerous [18,19]. Benefits include the accessibility to useful tools and resources for mental health diagnosis and treatment, like the one presented in this very paper. Risks involve very negative mental health outcomes such as an increase in suicidal ideation [20] and the exacerbation of mental illness stigma [21].

We think that digital ecosystems, like social media, connected videogames and instant messaging platforms, pose both opportunities and risks for mental health promotion. Therefore, it is the responsibility of policy makers, implementers and mental health professionals to pay real attention to this domain and promote a healthy use of online platforms. In a world where adolescents and young adults fail to express their psychological concerns to physicians but share them with the public in social media [22], we might need to, at least partially, move our prevention, evaluation and treatment operations to the online arena.

In that very line of action, we propose in this work the use of conversational agents to provide online platforms users with an engaging tool for depression screening. In the next section we describe the use of chatbots and conversational agents in healthcare, then, in Section 3, we outline the design and implementation of Perla, our agent. Section 4 covers the description of the validation study that we performed using both Perla and the PHQ-9 questionnaire. Finally, we offer a discussion of the results and the main conclusions drawn during our experience with Perla and the participants in the validation study.

## 2 Conversational Agents for Mental Health

The idea of using chatbots or automated conversational agents in mental health is not new at all. In fact, the first chatbot in history, the famous seminal work of Weizenbaum back in the 60's, called ELIZA [23], was supposed to act as a Rogerian therapist. Of course, it takes much more than a set of pre-programmed rules to become an accomplished psychotherapist, either human or artificial. It is our view that the very inflated expectations that were characteristic of the early ages of Artificial Intelligence (AI) has revived today in the context of current AI popularity.

Although current technological development in AI, Natural Language Understanding (NLU) [24], Deep Learning [25] and Conversational Interfaces [26,27,28] is astonishing, we think the idea of having a fully automated and capable psychotherapist is way too ambitious today, almost as it was back in the 60's. Actually, we consider this quest such a formidable challenge as that of passing the Turing Test [29]. However, as it is the case with the Turing Test, there are many limited and restricted forms of conversational challenges that we can define for artificial conversers. As a result of this, we can find today a plethora of different mental health solutions based on conversational agents [27,28]. In the context of this work, we distinguish the following types of systems:

- Mental health screening agents: designed to apply their conversational features in the early detection and psychological triage of different mental health problems. This is the case of the agent Perla presented in this work.

- Psychological evaluation agents: aimed at performing a thorough psychological evaluation as might be conducted in a psychological interview with a mental health professional.
- Psychological intervention agents: aimed at delivering specific psychological treatment through their conversational capabilities.
- Psychotherapist agents: aimed at potentially replacing human psychotherapists.

As discussed above, we think psychotherapist agents live only in the realm of science fiction. However, we also believe that remarkable advances can be currently achieved in the form of the other three types of agents. A recent example of a psychological intervention agent aimed at treating depression and anxiety is Woebot [30]. This system is based on delivering Cognitive Behavior Therapy (CBT) through a mobile application bot, and after a randomized controlled trial showed to be an effective way to reduce symptoms of depression and anxiety.

In general, we also believe that psychological intervention agents are useful to provide psychoeducational material to digital users in a more attractive and engaging way, thus making possible for mental health providers to take a step forward in the necessary move towards an active presence of mental health services in the digital ecosystem.

As part of our particular research effort, we feel that is important to assure a good quality, effectiveness and reliability in the design of mental health screening agents before progressing onto much more complex forms of conversational agents. For this reason, we developed Perla, as an interactive assessment tool for depression screening, and performed the corresponding validation study. Other mental health screening agents have been recently developed [31,32], addressing issues such as detecting suicide risk [33], depression [34], social needs [35], and several other mental illnesses [36], etc. However, most of the agents are purely experimental and lack high-quality evidence derived from randomized controlled studies [32]. Additionally, most of the recently developed agents are only available for English speaking users. Hence, we decided to make Perla a native Spanish speaker, so we can offer these depression screening services in Spain and Latin America.

## 3 Design and Implementation of Perla

Perla has been designed as a conversational agent able to perform a structured interview based on the PHQ-9 questionnaire. The main objective of Perla is to effectively estimate the presence of depression symptoms for Spanish speaking population. Perla conversation flow is based on the validated Spanish version of PHQ-9 [37]. One example of a question adapted from the English original version of PHQ-9 is "*Over the las 2 weeks, how often have you been bothered by feeling* *tired or having little energy?*" (please see below the complete set of questions in Spanish).

Perla's main features are:

- Natural language understanding: so, the user is not required to select a Likert option, but to naturally express how he or she feels in relation with each question (see Fig. 1).
- Text and Speech: verbal interaction with Perla is available both orally and written.
- Depression screening: Perla calculates the interviewee's PHQ-9 score and decide if it exceeds the cut-off point (PHQ-9 score $\geq$ 10).
- Feedback: Perla provides the participant with real-time feedback about the screening results and recommends psychoeducational material and links to mental health service when appropriate.
- Multiple online channels: Perla is designed to operate and be integrated in major popular platforms such as Skype, Hangout, Facebook Messenger, Google Assistant, Twitter, Telegram, telephone, etc.
- Data protection: screening results are stored in a real-time database once they are anonymized.
- Third-Party APIs: Perla is designed to be integrated with third-party Application Program Interfaces, so data can be shared in real-time to generate screening campaign results (current implementation generates a scientific report for the validation study).

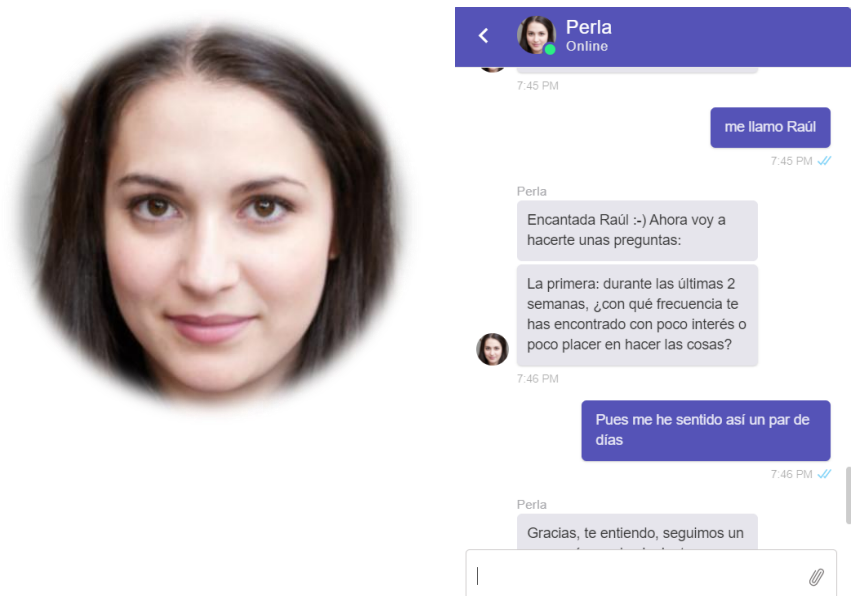

**Fig. 1.** Perla's human-like synthetic face (left) and instant messaging interface (right).

Perla has been implemented using the following components: Google DialogFlow [38], as the main NLU and conversation flow engine, Google Cloud Functions as serverless backend fulfillment [40], Firebase real-time database [39] for persisting data across stateless interactions, and Kommunicate [41] as the web chat interface used for the validation study. Fig. 2 shows a diagram of Perla's architecture.

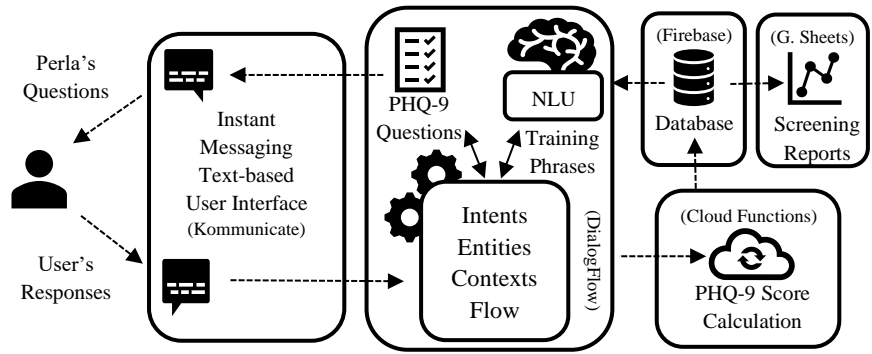

**Fig. 2.** Perla's architecture as implemented for the validation study.

Perla conversational flow is designed in the form of integrated contexts, intents and entities [26]:

- Contexts are used to control the flow of conversation and keeping track of the entities and parameters that are related with current semantic content. In our case, we use contexts to orchestrate the specific structure of the depression screening interview.
- Intents are typically used to categorize end-user's intention, therefore in Perla we use them to recognize user's responses. For the main part of the conversation, which is based on the 9 items of the PHQ-9 questionnaire, user's responses are expected to be a natural language expression referring to how frequently the user is experiencing a given depression symptom. For instance, a common response might be "*Oh, that happens to me all the time!*". More than 200 training phrases have been used for each PHQ-9 question to effectively train the NLU deep learning algorithm that matches user's responses.
- Entities are designed to identify specific information conveyed in user's intents. In Perla, we defined a specific entity in order to translate natural responses into an item score, which is expected to be equivalent to the one obtained when a Likert scale is used. As the PHQ-9 questionnaire is based on 4 Likert levels, we defined 4 entity values, accordingly, ranging from 0 points (meaning "not at all") to 3 points (meaning "nearly every day"). Each entity level is defined with more than 100 synonym phrases and fuzzy matching mode (to allow the recognition of partial or misspelled words).

In Table 1 we summarize the questions adapted from the Spanish version of PHQ-

9 questionnaire [37] as included in Perla's conversational flow. One of the problems of adapting a questionnaire to a live interview-like conversation is that the user might request clarification. Although Perla is capable of rephrasing questions upon request, we decided, for the sake of validity, to keep just one accurate version for each question. Therefore, clarification requests are handled with a standard reply, triggered by a fallback follow-up intent, where Perla clarifies that she is asking how frequently the symptom is experienced by the participant.

**Table 1.** PHQ-9 items (Spanish version) as formulated by Perla conversational agent.

| PHQ-9 Items | Perla formulation in Spanish |
|---|---|
| Item 1 (PI1) | *Durante las últimas 2 semanas, ¿con qué frecuencia te has encontrado con poco interés o poco placer en hacer las cosas?* |
| Item 2 (PI2) | *¿Con qué frecuencia te has sentido decaído/a, deprimido/a o sin esperanzas?* |
| Item 3 (PI3) | *¿Con qué frecuencia has tenido problemas de sueño (dificultad para quedarte dormido/a o dormir demasiado)?* |
| Item 4 (PI4) | *¿Con qué frecuencia te has sentido cansado/a o con poca energía?* |
| Item 5 (PI5) | *¿Con qué frecuencia has estado sin apetito o has comido en exceso?* |
| Item 6 (PI6) | *¿Con qué frecuencia te has sentido mal contigo mismo/a, que eres un fracaso o que has quedado mal contigo mismo/a o tu familia?* |
| Item 7 (PI7) | *¿Con qué frecuencia has tenido dificultades para concentrarte en actividades como leer o ver la televisión?* |
| Item 8 (PI8) | *¿Con qué frecuencia te has movido muy lento o has estado inquieto/a y agitado/a, moviéndote más de lo normal?* |
| Item 9 (PI9) | *¿Con qué frecuencia has pensado que estarías mejor muerto o en hacerte daño de alguna manera?* |

In the next section, we present the design and results of the validation study that we conducted in order to verify that Perla is a valid and reliable tool for depression screening.

## 4 Perla Validation Study

### 4.1 Participants

Following the philosophy of the application of Perla in digital ecosystems, a group of 276 adult Spanish-speaking participants were recruited directly through a social media campaign (see Materials and Procedures). The validation group was formed by 108 participants (39.13% of the global group; 65.7% female and 34.3% male;

normal distribution of age with an average of 37.21 years and 9.94 standard deviation) who also volunteered for depression screening using the standard self-report PHQ-9 questionnaire. We used this last group, who took both tests, to calculate validity and reliability indicators.

## 4.2 Materials and Procedures

The social media campaign to recruit participants consisted in a simple announcement posted simultaneously in Facebook, LinkedIn, Twitter and Instagram (social media advertisement was discarded so we could obtain an unbiased measure of user's interest). Perla's announcement post was actively accepting participants for two weeks and consisted of:

- A simple claim (*"Perla is an Artificial Intelligence that helps diagnose depression. Talk to her to learn how she does it!"*).
- The link to both Perla's chat-based interface and PHQ-9 form.
- The following hashtags: #chatbots, #psychology and #artificialintelligence.
- A picture of a female young lady, allegedly Perla, that was generated by a generative adversarial network (StyleGAN2) [42] (see Fig. 1).

Participants were recruited using a landing page in which they were invited to chat with Perla. After accepting the informed consent, the interview was performed automatically by Perla, who also stored screening results and contact data. All participants were also asked to fill out an ad hoc electronic version of the Spanish adapted version of the PHQ-9 questionnaire [37]. All the participants that took the interview with Perla but didn't take the standard PHQ-9 test were sent a reminder by email, two days after their first interaction, asking them to complete the PHQ-9 questionnaire. If, after 10 days, they failed to fill out the PHQ-9 questionnaire, a final reminder was sent. Despite these reminders, only 39.13% of the initially recruited participants completed the standard form.

Both assessment tools, Perla and PHQ-9 electronic form, were programmed to anonymously store the scores associated with each item as well as the total depression score. Additionally, the standard PHQ-9 cutoff point (score $\geq$ 10) was applied in order to calculate a class for each participant (either negative or positive). In the following, we use the nomenclature [*I1, I2, I3, ..., PHQ-9*] to refer to scores obtained from PHQ-9 form items and [*PI1, PI2, PI3, ..., PPHQ9*] to refer to the corresponding scores obtained by Perla.

## 4.3 Results

In relation with the participant recruitment process, while other content posted from the same social media account (the author's) has an average reach of approximately 4000 views in two weeks, Perla's validation study post reached 11,266 visualizations in the same period (2.82 times more reach) and was shared by other 32 social media accounts. The participation ratio for Perla screening was 2.5%, while the traditional PHQ-9 questionnaire (electronic form) was only filled out by 0.96% of reached users.

The number of depression cases detected by Perla and the PHQ-9 form were slightly different: according to Perla 28.70% of participants are in high risk of suffering a depression-related disorder, while data from the PHQ-9 questionnaire indicates a prevalence of 22.23% (this unusually high prevalence of depression for a non-clinical population is explained by the effects of SARS-CoV2 pandemic, see Discussion section below). Despite the distinct prevalence ratios reported by the two methods, there are not statistically significant differences (one-way ANOVA f=1.191; $p > 0.05$) and the two screening methods are highly correlated – see Fig 3 (Pearson's correlation of 0.91 for PHQ-9 score and 0.79 point-biserial correlation for dichotomic depression screening results).

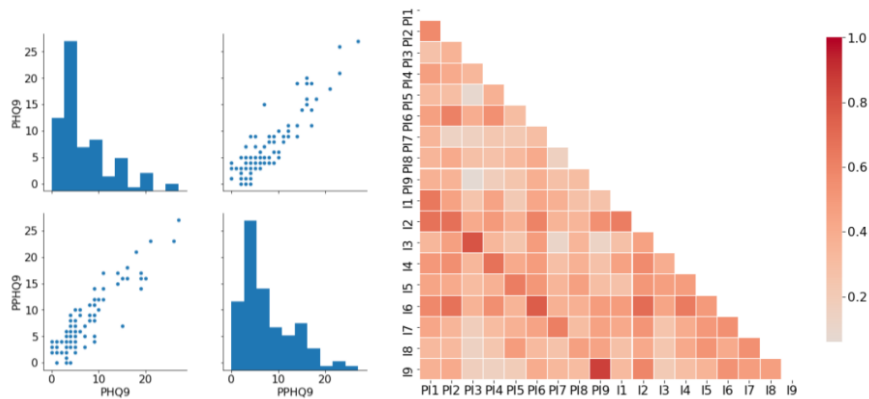

**Fig. 3.** Correlation plot for PHQ-9 score (left) and items score correlation (right).

As shown in Fig. 4, both screening tools yield the same distribution differing by a mean absolute error (MAE) of 1.88 points (1.61 standard deviation). We didn't find any significant correlation between the MAE and the number of days separating the two measures, which ranged from 0 to 14 days.

Given the unbalanced nature of the score results dataset (where negative cases of depression constitute the great majority), we analyzed the screening capability of Perla considering her a classifier, and therefore applying signal detection theory [43]. From this point of view, we have initially calculated the inter-rater reliability [44], considering Perla and the PHQ-9 form as two different raters for the presence of depression. The inter-rater reliability evaluation results in a Cohen's kappa of 0.77 for depression classification (0.20 for the specific PHQ-9 score), which indicates a substantial agreement according to Landis and Koch's interpretation [45]. Specific statistics for each item are presented in Table 2.

**Table 2.** Pearson's correlation coefficients (PCC), inter-rater reliability (kappa), accuracy (ACC) and mean absolute error (MAE) for all items of Perla/PHQ-9 questionnaire.

| | I1 PI1 | I2 PI2 | I3 PI3 | I4 PI4 | I5 PI5 | I6 PI6 | I7 PI7 | I8 PI8 | I9 PI9 | PHQ9 PPHQ9 |
|---|---|---|---|---|---|---|---|---|---|---|
| PCC | 0.65 | 0.68 | 0.79 | 0.67 | 0.63 | 0.76 | 0.62 | 0.47 | 0.86 | **0.91** |
| kappa | 0.45 | 0.41 | 0.47 | 0.44 | 0.44 | 0.56 | 0.36 | 0.31 | **0.72** | 0.20 |
| ACC | 0.62 | 0.61 | 0.62 | 0.61 | 0.66 | 0.73 | 0.57 | 0.66 | 0.91 | **0.92** |
| MAE | 0.49 | 0.45 | 0.42 | 0.47 | 0.47 | 0.32 | 0.53 | 0.49 | 0.09 | **1.88** |

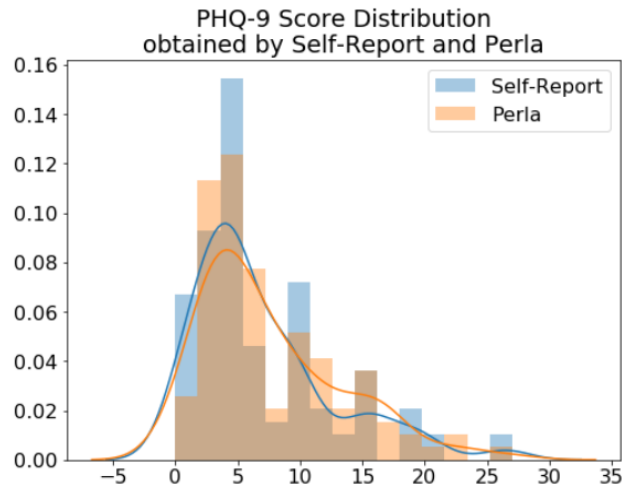

**Fig. 4.** Comparison of PHQ-9 score histograms (Perla vs electronic form).

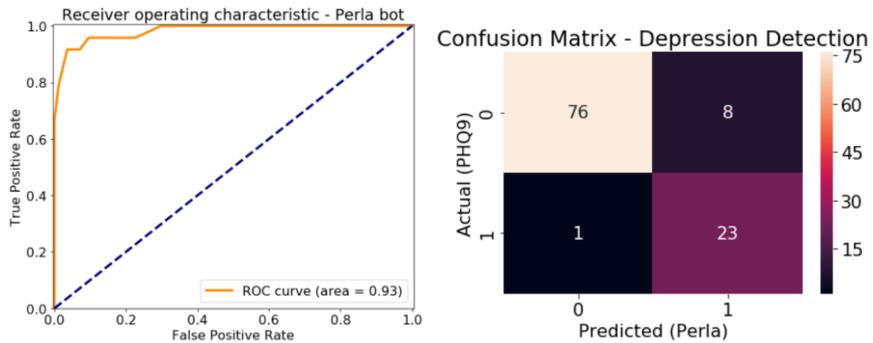

**Fig. 5.** Perla's Receiver Operating Characteristic (ROC) curve and confusion matrix.

As a depression detector, taking the PHQ-9 form results as ground truth, Perla behaves as presented in Fig. 5, with an area under the ROC curve (AUC) of 0.93, a sensitivity of 0.96, specificity of 0.90, accuracy of 0.92 and F1-score of 0.84.

In relation with the reliability of Perla as depression screening tool, we have calculated Cronbach's α [46], obtaining a value of 0.81, which indicates a very good consistency, though inferior to the alpha obtained by the electronic PHQ-9 form used in our study (0.88), which, in fact, parallels the excellent value reported originally by the authors of the scale (0.89) [11].

### 4.4 Discussion

One of the most remarkable insights that we have discovered working with Perla is that social media users are much keener to participate in a psychological screening if the task involves interacting with an AI-based agent. Even though completing the standard 9-items form is much faster, Perla has been more than 2.5 times more popular.

A specific survey might be required in order to find out why exactly users prefer to talk to a conversational agent. It is out hypothesis that this general preference is a combination of factors, involving the current popularity of AI applications, curiosity and the desire to express oneself beyond the restrictive pattern of a Likert scale. In general, digital users seem to be inclined to use and early adopt new forms of interaction with artificial agents. We also think that giving the agent some anthropomorphic features, like having a human face and name, makes the experience more appealing and facilitates engagement thanks to personification processes [47].

As any classifier, Perla has a specific bias in the balance between false positives and false negatives. As can be observed in the confusion matrix (Fig. 5), Perla is inclined to err by generating more false positives that false negatives (high *d'*, hence high sensitivity). This effect is indeed produced by design, given the way we match the entity responses with the corresponding item score, and we think it is a good feature for a screening or triage tool.

More investigation and more complex research designs might be required in order to find out whether some of Perla's false positives are in fact 'hidden' cases of depression not correctly detected by the original PHQ-9 instrument. Our untested hypothesis here is that moving from a Likert scale framework to a structured (automated) interview context makes it possible for the users to express more subtleties and details about their feelings. We think that the entity recognition mechanism that Perla applies to natural language responses has contributed to higher depression risk scores. Interestingly, Perla's scores are only higher above the PHQ-9 cutoff point (see Fig 4).

The validation group recruited for this study seems to provide a statistically significant random sample, however, there are obvious biases, like the unbalanced gender distribution (nearly two-third female subjects), which does not match the reference population. In order to have a stronger evidence of the validity of Perla a much bigger and representative group would need to be assembled for testing. That would also imply to design a new recruitment and/or participant selection procedure.

In relation to the high proportion of positives detected by both Perla and the standard PHQ-9 form, which does not correspond to typical depression prevalence for Spanish population (that should be, in normal circumstances, around 10%) [48,49,50], we think the cause is the acute current situation of the SARS-CoV2 pandemic crisis, that raised Spanish prevalence of depression up to 34% during the recent lockdown [51,52,53], and now seems to be decreasing progressively.

## 5 Conclusions

The results obtained during the validation study show that using a conversational agent for depression screening is a valid alternative to traditional self-report tools. Transforming the Likert scale into a structured interview conducted by an artificial agent do not imply a significant loss of reliability and increases the acceptance and engagement of online users. In light of the above, a tool like Perla can be considered a valuable asset in the promotion of mental health in the online world.

Finding a low cost, appealing and effective way to expand the early diagnosis of depression to the greater part of the population is a challenge, but also a key factor contributing to timely treatment and better prognosis. We have conceived Perla as an AI-based resource that leverages the power of digital ecosystems, like social media and online communities, where an ever-increasing segment of the general population can be reached. Furthermore, agent-based screening might contribute to the early detection of depression in those reluctant to seek out help outside cyberspace.

The feedback obtained from the participants of this study makes us think that bringing natural language interfaces to the world of psychometrics and digital health is a great opportunity with a lot of potential. The automatization of natural interaction makes it possible to go beyond self-report tests without the huge cost involved in conducting interviews by trained human professionals. In this regard, we see that the clinical interview is a much more powerful way of evaluation than self-report instruments [54,55], but for screening purposes we typically must conform with the self-report option, given that conducting interviews is not a scalable choice. The application of conversational interfaces, not only for well-structured interviews, like we have implemented in Perla, but also semi-structured interviews, has the potential to effectively extend mental health screening and preven-tion to a much wider population.

Of course, we do not advocate the substitution of mental health professionals with machines, as we do not even consider the possibility of the "psychotherapist agent" type. However, current advances in machine vision and sensor pattern recognition might be of great advantage when combined with natural language understanding capabilities. This might lead to the design of "psychological evaluation agents" able to incorporate non-verbal communication clues into their assessment process.